\documentclass{sigchi-ext}
\usepackage[T1]{fontenc}
\usepackage{textcomp}
\usepackage[scaled=.92]{helvet} 
\usepackage{graphicx} 
\usepackage{balance}  
\usepackage{booktabs} 
\usepackage{ccicons}  
\usepackage{ragged2e} 

\def\plaintitle{An implementation of an imitation game with ASD children to learn nursery rhymes : first tests} 
\def\emptyauthor{}
\def\plainkeywords{imitation; socially assistive robotics; Pepper; openpose; pose estimation; serious game; autism}

\setlist{leftmargin=1.mm}

\title{An implementation of an imitation game with ASD children to learn nursery rhymes}

\numberofauthors{6}
\author{%
  \alignauthor{%
    \textbf{Sao Mai Nguyen}\\
    \affaddr{IMT Atlantique} \\
    \affaddr{Brest, France} \\
    \small{\email{nguyensmai@gmail.com}} 
  } 
   \alignauthor{%
    \textbf{Sebastien Guillon\\
    Patricio Tula\\
    Alvaro Paez\\
    Mouad Bouaida\\
    Arthus Anin\\
    Saad El Qacemi}\\
    \affaddr{IMT Atlantique}\\
    \affaddr{Brest, France}\\
  }\\[-8ex]
  \alignauthor{%
    \textbf{Nathalie Collot-Lavenne}\\
    \affaddr{Brest University Hospital}\\
    \affaddr{Bohars, France}\\
    \small{\email{nathalie.lavenne-collot@chu-brest.fr}} 
  }\\[2ex]
  \alignauthor{%
    \textbf{Christophe Lohr}\\
    \affaddr{IMT Atlantique}\\
    \affaddr{Brest, France}\\
    \small{\email{christophe.lohr@imt-atlantique.fr}} 
  }
\\[2ex]
  \alignauthor{%
The authors would like to thank Celine Degrez-Foudrat and Camille Mallegol for their participation in this work. 
  }
 }

\definecolor{linkColor}{RGB}{6,125,233}
\hypersetup{%
  pdftitle={\plaintitle},
  pdfauthor={\emptyauthor},
  pdfkeywords={\plainkeywords},
  bookmarksnumbered,
  pdfstartview={FitH},
  colorlinks,
  citecolor=black,
  filecolor=black,
  linkcolor=black,
  urlcolor=linkColor,
  breaklinks=true,
}

\begin{document}


\maketitle

\RaggedRight{} 

\begin{abstract}

Previous studies have suggested that being imitated by an adult is an effective intervention with children with autism and developmental delay. The purpose of this study is to investigate if an imitation game with a robot can arise interest from children and constitute an effective tool to be used in clinical activities. In this paper, we describe the design of our nursery rhyme imitation game, its implementation based on RGB image pose recognition and the preliminary tests we performed.
 \end{abstract}


\category{See}{\url{http://acm.org/about/class/1998/}}{for
  full list of ACM classifiers. This section is required.}

\section{Introduction}

Several studies have shown that imitating ASD children can be efficient in enhancing their social behaviour \cite{Nadel2005Imitationetaustisme}. These studies have been carried with caregivers as imitators of children's movements \cite{Sanefuji2009IMHJOPWAIMH,Katagiri2010RASD}.  Upon these results, numerous projects have been using TIC including robots, for ASD care \cite{Sartorato2017JPR,Huijnen2017JADD}. For instance, project Gaming Open Library for Intervention in Autism at Home (GOLIAH) developed a series of 11 serious games on computers and tablets using imitation and joint attention \cite{Jouen2017CAPMH}.  Other projects have focused on facial expression with robot Keepon \cite{Kozima2009IJSR} or with the expressive robot Face \cite{Pioggia2007R21IISRHIC}. For instance, Kaspar \cite{Dautenhahn2009ABB} can be used with the capacity to imitate drum rhythm.

\begin{marginfigure}[0pc]
  \begin{minipage}{\marginparwidth}
    \centering
    \includegraphics[width=0.9\marginparwidth]{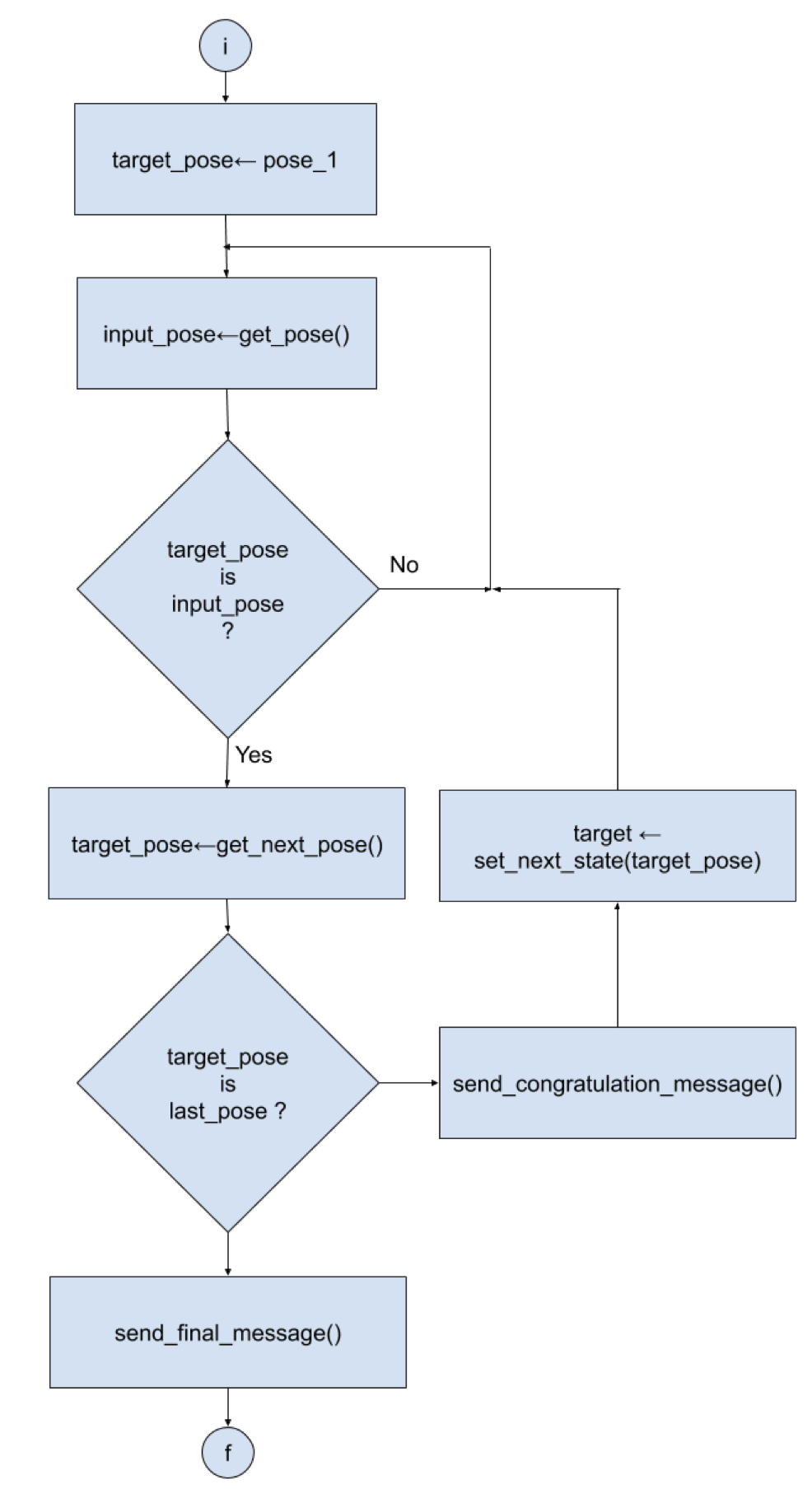}
    \caption{Diagram of the flow of the processes of the game}\label{fig:diagramFlux}
  \end{minipage}
\end{marginfigure}

\begin{marginfigure}[2pc]
  \begin{minipage}{\marginparwidth}
    \centering
  \includegraphics[width=0.9\columnwidth]{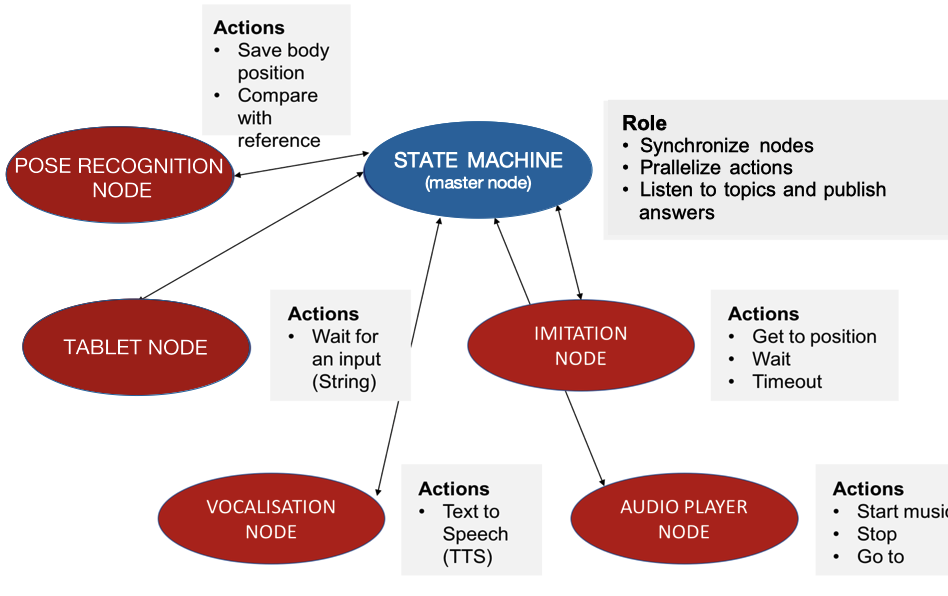}
  \caption{Structure of the ROS nodes.}\label{fig:rosNodes}
  \end{minipage}
\end{marginfigure}

While most imitation studies have focused on facial expressions or sounds, few studies have focused on the imitation of movements. However, impairment in motor imitation abilities have been commonly described in children with ASD. How motricity in interpersonal coordination impacts imitation has little been investigated. Indeed, \cite{Xavier2018FP} has shown that both interpersonal synchronization and motor coordination subtly contribute the imitation impairment in ASD children. A human partner was shown to have an impact on the learning abilities of postures by a humanoid robot \cite{Guedjou20162JIICDLERI}, and when it was interacting with participants, it was able to learn a social signature  at the level of individual recognition \cite{Boucenna2016SR}. These results are interesting, but are only limited to static postures and not yet dynamic movement as required in full motor control. 
Development of motor control requires forming an internal model of action relying on the coupling between action (motor commands) and perception (sensory feedback). Critical to the development of social, communicative, and motor coordination behaviors, internal model of action accurately predicts the sensory consequences of motor commands \cite{Krakauer2007CNTIIBF}. This is why in this study, we propose to examine how ASD children can learn to improve their motor control by a serious game. We propose to use a robot platform as they have been shown to arise their interest.

To allow them to learn motor control and gestures, we examined the program of clinical activities and identified that during their activities, nursing rhymes were sung and mimicked with therapists. We therefore propose to implement a robot for an imitation game based on a nursery rhyme. We describe our implementation and the test results.

\section{General architecture of our system}
We implemented a software for the humanoid robot Pepper to play an imitation game with a nursing rhyme. The robot would thus sing a line of the rhyme while making a gesture and displaying an image on its table, then check if the child imitates the gesture, while encouraging him and displaying an image of the nursery rhyme. This workflow is illustrated by fig. \ref{fig:diagramFlux}.

The system is composed of: (i) pose recognition node; (ii) a display node using Pepper's tablet;  (iii) a node to control the Pepper robot movements using Naoqi \cite{naoqi} through the ROS module MoveIt \cite{Chitta2012IRAM}; (iv) the audio player node to play the music line by line (v) the vocalisation node for text-to-speech.  These nodes can communicate through ROS as shown in fig. \ref{fig:rosNodes}. The role of the master node is to synchronise all nodes and to manage the behaviour of the robot.

\section{Pose recognition}

\begin{marginfigure}[-25pc]
  \begin{minipage}{\marginparwidth}
    \centering
    \includegraphics[width=0.9\marginparwidth]{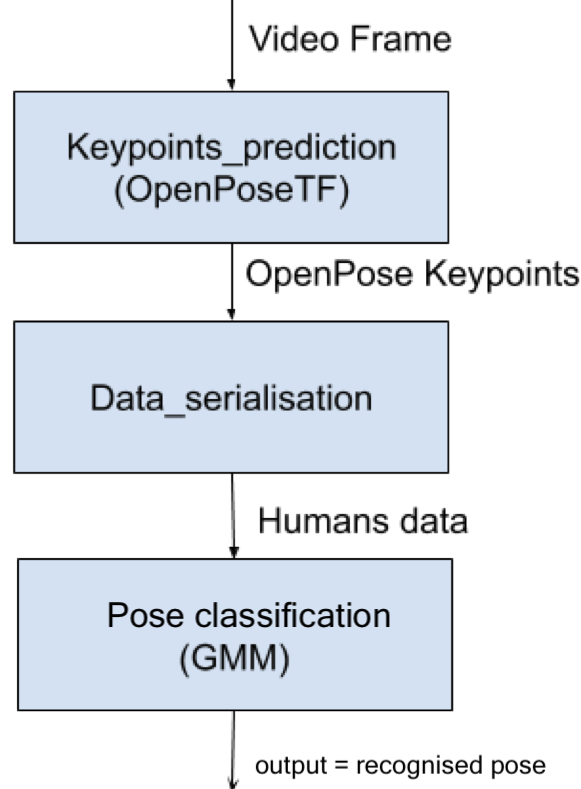}
    \caption{Architecture of the different modules to recognise poses and control the robot}\label{fig:systemArchitecture}
  \end{minipage}
\end{marginfigure}
\begin{marginfigure}[-4pc]
  \begin{minipage}{\marginparwidth}
    \centering
    \includegraphics[width=0.9\columnwidth]{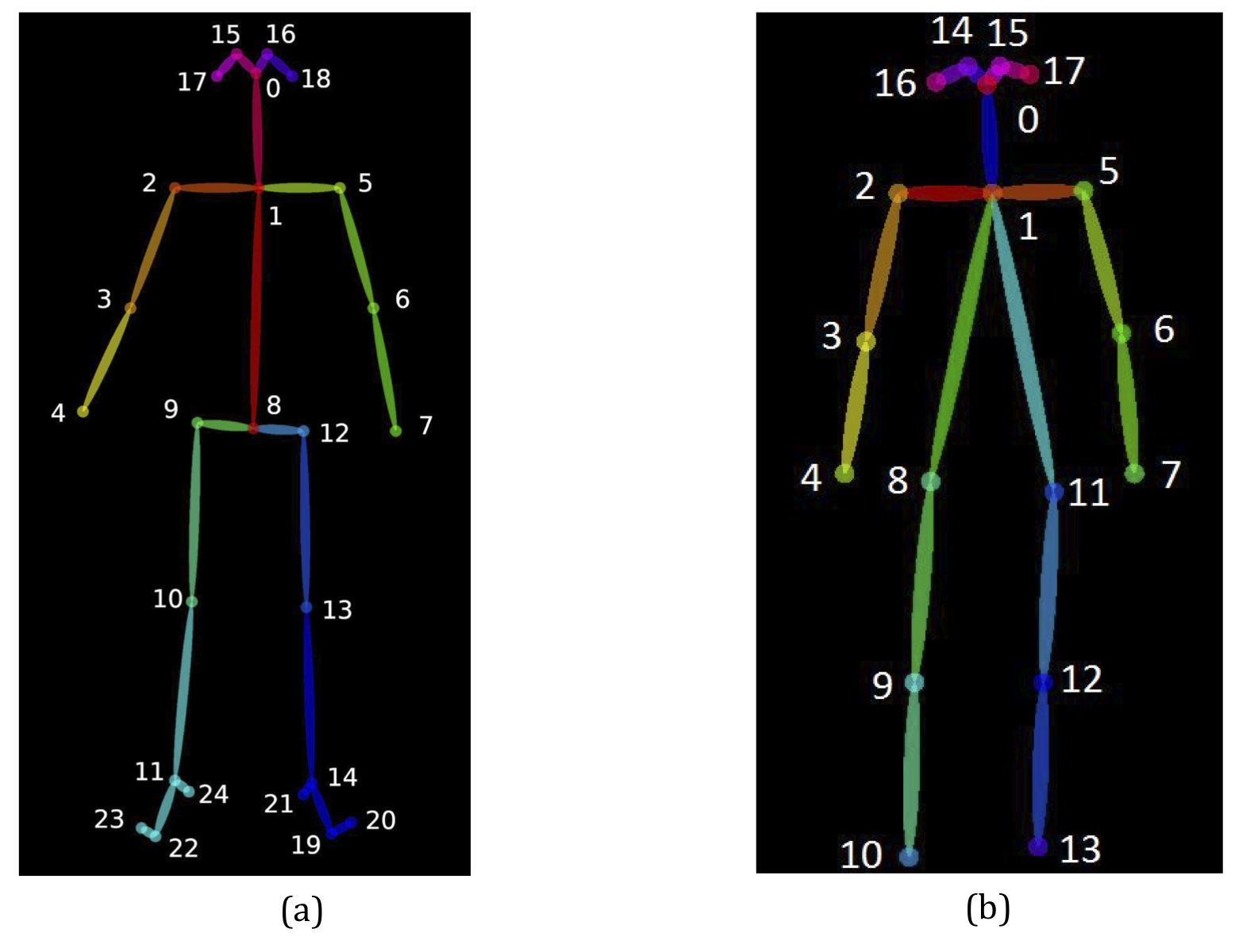}
  \caption{Openpose can track human pose "skeletons" using either (a) body-25 or (b) coco model.}\label{fig:skeleton}
  \end{minipage}
\end{marginfigure}

Our pose recognition system (fig. \ref{fig:systemArchitecture}),  is composed of  :
\begin{itemize}
\item the internal RGB camera of Pepper giving to the system the input video flow.
\item the OpenPose library \cite{Cao2017C} to estimate the pose of the human body using deep learning algorithm, and a classifier to recognise a predefined set of poses. For our application, we used the tensorflow implementation on CPU of openpose, instead of the more vanilla implementation for GPU. The openpose library can output two models of skeleton with either 25 or 18 joints called body-25 and coco, as illustrated in fig. \ref{fig:skeleton}. In our case, as the nursery rhyme did not focus on the bottom body part, we opted for the coco model. To avoid false positives, we set the confidence threshold c to 0.5. The openpose library returns at each frame the skeleton features $((x_1,y_1) , ..., (x_8,y_8))$ of the positions (x,y) of the 8 upper body joints  (numbered 1 to 8) if its confidences on these joints are above c. These skeleton features are normalised as
\begin{equation}
input\_pose_x = \frac{x - x_{0}}{|x_{1} -x_{2}|}~, ~
input\_pose_y = \frac{y - y_{0}}{|y_{3} -y_{4}|}
\end{equation}
where $(x_0,y_0)$ is the intersection of joints 1-2 and 3-4.
\item a classifier:  we used  Gaussian Mixture Classifier (GMM) with a dataset of 20 samples for each of the 8 class shown in fig. \ref{fig:rhymePoses}. In the work presented here, the GMM is used to classify poses, but GMM can also be used to classify dynamical movements, as well as evaluate an error in the movement so as to give an advice how to improve the movement, as described in \cite{Devanne2017IICHRH} in a project of a robot coach for physical rehabilitation.
\end{itemize}

\section{Evaluation}
We report here the performance of the pose recognition system, and describe our test with an autistic child.

\subsection{Performance of the pose recognition system}
\begin{marginfigure}[-0pc]
  \begin{minipage}{\marginparwidth}
    \centering
    \includegraphics[width=0.9\marginparwidth]{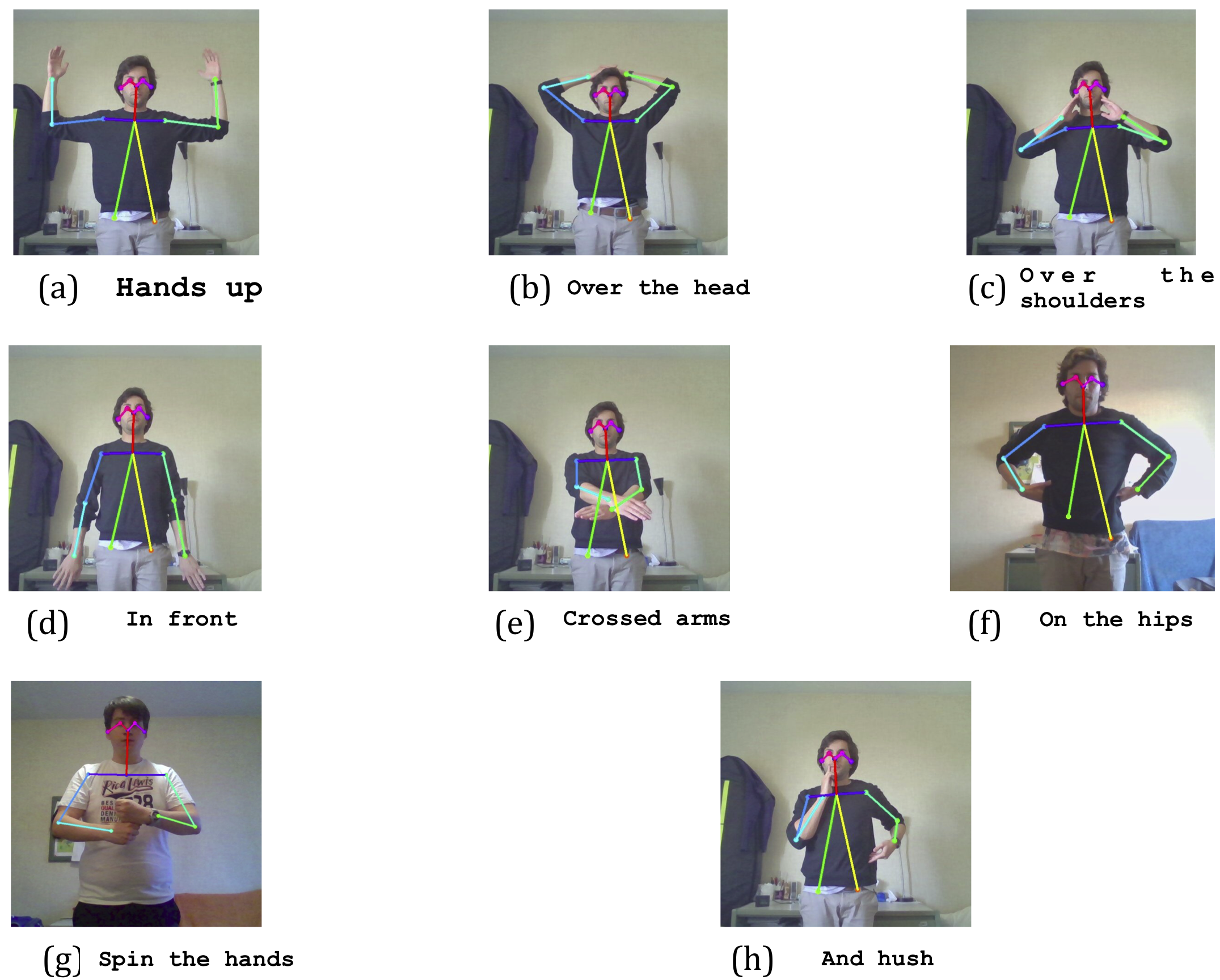}
    \caption{Poses used for the nursery rhyme and the skeletons detected by openpose.}\label{fig:rhymePoses}
  \end{minipage}
\end{marginfigure}

We performed an evaluation on the performance of our pose recognition system on the 8 classes of poses in the rhyme. They are shown in fig \ref{fig:rhymePoses}. The image of the database were captured by a webcam with a resolution of 640x480. We used for openpose the model CMU as it gives the best detects skeletons. Each class had 30 images. The train accuracy is 94.79\% while the test accuracy is 94.59 \%.

%

\subsection{Test with a subject}

We tested the idea of a nursery rhyme game with an autistic children before fully implementing our pose recognition system. For this test, we used a wizard of oz. An engineer was remote-controlling the robot. We tested in 2 configurations: the first when the robot is static, and the second when the robot can move around. In both configurations, the robot can sing the rhymes and make the gestures of the rhymes, and wait for the child to perform.
Our subject is a 4-year-old male who shows typical characteristics of infant autism, featuring alterations of social communication. He also shows limited, repetitive and stereotypical interest and behaviours, conform to the diagnostic criteria DSM V and APA. He shows an important developmental delay in language, can only say a few isolated words, mostly as an echo, and avoids eye contact.

We noticed a clear interest of the subject in the robot and his interest to interact at the beginning of both sessions. However, we noted a clear improvement of the attention in the case the robot can move, and regain attention, whereas in the case of a static robot, the child is easily distracted by other details of the room.

\section{Discussion}
These tests show interest of subjects in our proposed imitation game. However, we still need to improve our system. First, we need to extend our algorithm to recognise not only poses, but also dynamic movements such as in \cite{Devanne2017IICHRH}. Then, we would need to test our pose recognition system on our subjects, and implement not only a control system of the arms and voice for the nursery rhyme but also a control of the position of the robot to be able to attract the attention of the subject. Moreover, our system needs a good interaction system so as to know when to wait for the child to mimic and sing along the rhyme, or to repeat the line of the rhyme to remind the child of the rhyme, or even to continue to the next line to sustain his interest. The implementation of such an autonomous robot would require a difficult understanding of the behaviour of the subject to detect not only if he has repeated the gestures, but also whether the absence of response is due to a lack of time to respond, a bad understanding of the gesture, a lack of attention, or a lack of interest.

\begin{marginfigure}[10pc]
  \begin{minipage}{\marginparwidth}
    \centering
    \includegraphics[width=0.9\marginparwidth]{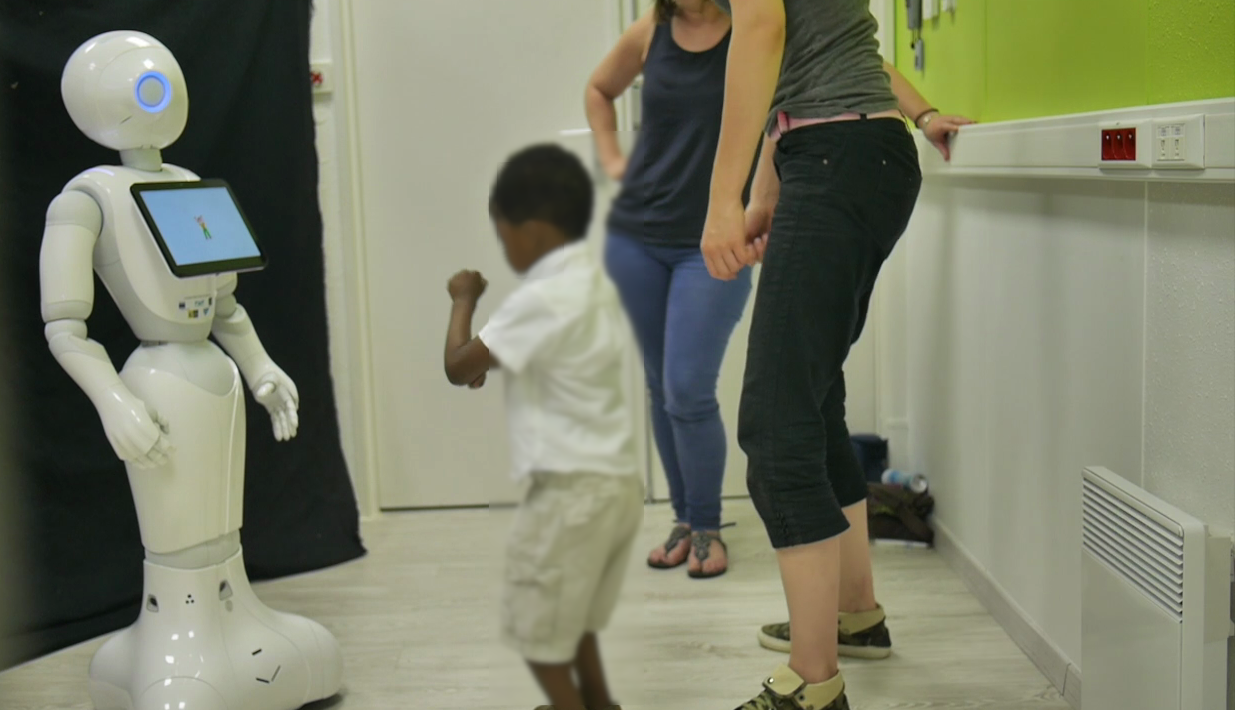}
    \caption{During the test session with the mobile robot: the subject is making the gesture "Spin the hands" in the imitation game}\label{fig:subjectTestBlur}
  \end{minipage}
\end{marginfigure}

These first tests support a use of a humanoid robot as a therapeutic tool to improve imitation skills in ASD children. Indeed, imitation constitutes a pivotal skill in children to enable the development of other cognitive and communication skills. 

\balance{} 
\footnotesize
\bibliographystyle{SIGCHI-Reference-Format}
\bibliography{main.bib}


\begin{thebibliography}{00}


\ifx \showCODEN    \undefined \def \showCODEN     #1{\unskip}     \fi
\ifx \showDOI      \undefined \def \showDOI       #1{{\tt DOI:}\penalty0{#1}\ }
  \fi
\ifx \showISBNx    \undefined \def \showISBNx     #1{\unskip}     \fi
\ifx \showISBNxiii \undefined \def \showISBNxiii  #1{\unskip}     \fi
\ifx \showISSN     \undefined \def \showISSN      #1{\unskip}     \fi
\ifx \showLCCN     \undefined \def \showLCCN      #1{\unskip}     \fi
\ifx \shownote     \undefined \def \shownote      #1{#1}          \fi
\ifx \showarticletitle \undefined \def \showarticletitle #1{#1}   \fi
\ifx \showURL      \undefined \def \showURL       #1{#1}          \fi

\bibitem{naoqi}
 2019.
\newblock   (2019).
\newblock
\showURL{%
\url{http://doc.aldebaran.com/2-1/naoqi/index.html}}


\bibitem{Boucenna2016SR}
{Sofiane Boucenna}, {David Cohen}, {Andrew~N. Meltzoff}, {Philippe Gaussier},
  {and} {Mohamed Chetouani}. 2016.
\newblock \showarticletitle{Robots Learn to Recognize Individuals from
  Imitative Encounters with People and Avatars}.
\newblock {\em Scientific Reports\/}  {6} (04 02 2016), 19908 EP --.
\newblock
\showURL{%
\url{https://doi.org/10.1038/srep19908}}


\bibitem{Cao2017C}
{Zhe Cao}, {Tomas Simon}, {Shih-En Wei}, {and} {Yaser Sheikh}. 2017.
\newblock \showarticletitle{Realtime Multi-Person 2D Pose Estimation using Part
  Affinity Fields}. In {\em CVPR}.
\newblock


\bibitem{Chitta2012IRAM}
{Sachin Chitta}, {Ioan Sucan}, {and} {Steve Cousins}. 2012.
\newblock \showarticletitle{Moveit![ros topics]}.
\newblock {\em IEEE Robotics \& Automation Magazine\/} {19}, 1 (2012), 18--19.
\newblock


\bibitem{Dautenhahn2009ABB}
{Kerstin Dautenhahn}, {Chrystopher~L Nehaniv}, {Michael~L Walters}, {Ben
  Robins}, {Hatice Kose-Bagci}, {N~Assif Mirza}, {and} {Mike Blow}. 2009.
\newblock \showarticletitle{KASPAR--a minimally expressive humanoid robot for
  human--robot interaction research}.
\newblock {\em Applied Bionics and Biomechanics\/} {6}, 3-4 (2009), 369--397.
\newblock


\bibitem{Devanne2017IICHRH}
{Maxime Devanne} {and} {Sao~Mai Nguyen}. 2017.
\newblock \showarticletitle{Multi-level Motion Analysis for Physical Exercises
  Assessment in Kinaesthetic Rehabilitation}. In {\em IEEE International
  Conference on Humanoid Robots (Humanoids)}.
\newblock


\bibitem{Guedjou20162JIICDLERI}
{H. {Guedjou}}, {S. {Boucenna}}, {and} {M. {Chetouani}}. 2016.
\newblock \showarticletitle{Posture recognition analysis during human-robot
  imitation learning}. In {\em 2016 Joint IEEE International Conference on
  Development and Learning and Epigenetic Robotics (ICDL-EpiRob)}. 193--194.
\newblock


\bibitem{Huijnen2017JADD}
{Claire~AGJ Huijnen}, {Monique~AS Lexis}, {Rianne Jansens}, {and} {Luc~P de
  Witte}. 2017.
\newblock \showarticletitle{How to Implement Robots in Interventions for
  Children with Autism? A Co-creation Study Involving People with Autism,
  Parents and Professionals}.
\newblock {\em Journal of autism and developmental disorders\/} {47}, 10
  (2017), 3079--3096.
\newblock


\bibitem{Jouen2017CAPMH}
{Anne-Lise Jouen}, {Antonio Narzisi}, {Jean Xavier}, {Elodie Tilmont}, {Nicolas
  Bodeau}, {Valentina Bono}, {Nabila Ketem-Premel}, {Salvatore Anzalone},
  {Koushik Maharatna}, {Mohamed Chetouani}, {Filippo Muratori}, {and} {David
  Cohen}. 2017.
\newblock \showarticletitle{GOLIAH (Gaming Open Library for Intervention in
  Autism at Home): a 6-month single blind matched controlled exploratory
  study}.
\newblock {\em Child and Adolescent Psychiatry and Mental Health\/} {11}, 1 (22
  Mar 2017), 17.
\newblock
\showISSN{1753-2000}


\bibitem{Katagiri2010RASD}
{Masatoshi Katagiri}, {Naoko Inada}, {and} {Yoko Kamio}. 2010.
\newblock \showarticletitle{Mirroring effect in 2- and 3-year-olds with autism
  spectrum disorder}.
\newblock {\em Research in Autism Spectrum Disorders\/} {4}, 3 (2010), 474 --
  478.
\newblock
\showISSN{1750-9467}


\bibitem{Kozima2009IJSR}
{Hideki Kozima}, {Marek~P Michalowski}, {and} {Cocoro Nakagawa}. 2009.
\newblock \showarticletitle{Keepon}.
\newblock {\em International Journal of Social Robotics\/} {1}, 1 (2009),
  3--18.
\newblock


\bibitem{Krakauer2007CNTIIBF}
{John~W. Krakauer} {and} {Reza Shadmehr}. 2007.
\newblock \showarticletitle{Towards a computational neuropsychology of action}.
\newblock In {\em Computational Neuroscience: Theoretical Insights into Brain
  Function}, {Paul Cisek}, {Trevor Drew}, {and} {John~F. Kalaska} (Eds.).
  Progress in Brain Research, Vol. 165. Elsevier, 383 -- 394.
\newblock
\showISSN{0079-6123}


\bibitem{Nadel2005Imitationetaustisme}
{Jacqueline Nadel}. 2005.
\newblock {\em L'autisme}.
\newblock Chapter Imitation et austisme.
\newblock


\bibitem{Pioggia2007R21IISRHIC}
{Giovanni Pioggia}, {ML Sica}, {Marcello Ferro}, {Roberta Igliozzi}, {Filippo
  Muratori}, {Arti Ahluwalia}, {and} {Danilo De~Rossi}. 2007.
\newblock \showarticletitle{Human-robot interaction in autism: FACE, an
  android-based social therapy}. In {\em RO-MAN 2007-the 16th IEEE
  international symposium on robot and human interactive communication}. IEEE,
  605--612.
\newblock


\bibitem{Sanefuji2009IMHJOPWAIMH}
{Wakako Sanefuji}, {Hiroshi Yamashita}, {and} {Hidehiro Ohgami}. 2009.
\newblock \showarticletitle{Shared minds: Effects of a mother's imitation of
  her child on the mother--child interaction}.
\newblock {\em Infant Mental Health Journal: Official Publication of The World
  Association for Infant Mental Health\/} {30}, 2 (2009), 145--157.
\newblock


\bibitem{Sartorato2017JPR}
{Felippe Sartorato}, {Leon Przybylowski}, {and} {Diana~K. Sarko}. 2017.
\newblock \showarticletitle{Improving therapeutic outcomes in autism spectrum
  disorders: Enhancing social communication and sensory processing through the
  use of interactive robots}.
\newblock {\em Journal of Psychiatric Research\/}  {90} (2017), 1 -- 11.
\newblock
\showISSN{0022-3956}


\bibitem{Xavier2018FP}
{Jean Xavier}, {Soizic Gauthier}, {David Cohen}, {Mohamed Zahoui}, {Mohamed
  Chetouani}, {Fran{\c c}ois Villa}, {Alain Berthoz}, {and} {Salvatore
  Anzalone}. 2018.
\newblock \showarticletitle{Interpersonal Synchronization, Motor Coordination,
  and Control Are Impaired During a Dynamic Imitation Task in Children With
  Autism Spectrum Disorder}.
\newblock {\em Frontiers in Psychology\/}  {9} (Sep 2018).
\newblock
\showISSN{1664-1078}


\end{thebibliography}

\end{document}